\documentclass[prl,aps,epsf,twocolumn,showpacs,superscriptaddress]{revtex4}

\begin{document}
%
% Defintions   
%
\renewcommand{\Re}{\operatorname{Re}}
\renewcommand{\Im}{\operatorname{Im}}
\newcommand{\Tr}{\operatorname{Tr}}
\newcommand{\sign}{\operatorname{sign}}
\newcommand{\dd}{\text{d}}
\newcommand{\q}{\boldsymbol q}
\newcommand{\p}{\boldsymbol p}
\newcommand{\rr}{\boldsymbol r}
\newcommand{\pp}{p_v}
\newcommand{\vv}{\boldsymbol v}
\newcommand{\I}{{\rm i}}
\newcommand{\pphi}{\boldsymbol \phi}
\newcommand{\ds}{\displaystyle}
\newcommand{\be}{\begin{equation}}
\newcommand{\ee}{\end{equation}}
\newcommand{\bea}{\begin{eqnarray}}
\newcommand{\eea}{\end{eqnarray}}
\newcommand{\Acl}{{\cal A}}
\newcommand{\Rcl}{{\cal R}}
\newcommand{\Tcl}{{\cal T}}
\newcommand{\Tmin}{{T_{\rm min}}}
\newcommand{\Toff}{{\langle \delta T \rangle_{\rm off} }}
\newcommand{\Roff}{{\langle \delta R \rangle_{\rm off} }}
\newcommand{\RoffI}{{\langle \delta R_I \rangle_{\rm off} }}
\newcommand{\RoffII}{{\langle \delta R_{II} \rangle_{\rm off} }}
\newcommand{\dg}{{\langle \delta g \rangle_{\rm off} }}
\newcommand{\rd}{{\rm d}}
\newcommand{\br}{{\bf r}}
\newcommand{\la}{\langle}
\newcommand{\ra}{\rangle}

%
% \draft command makes pacs numbers print
%

 \title{Statistical description of eigenfunctions \\ 
 in chaotic and weakly disordered systems beyond universality}

\author{Juan Diego Urbina}
\affiliation{Institute for Physics of Complex Systems, The Weizmann Institute of Science,76100 Rehovot, Israel}

\author{Klaus Richter}
\affiliation{ Institut f{\"u}r Theoretische Physik, Universit{\"a}t Regensburg, 
         93040 Regensburg, Germany}

\begin{abstract}
We present a semiclassical approach to eigenfunction statistics in chaotic 
and weakly disordered quantum systems which goes beyond Random Matrix Theory,
supersymmetry techniques, and existing semiclassical methods. The approach is based 
on a generalization 
of Berry's Random Wave Model, combined with a consistent semiclassical representation
of spatial two-point correlations. We derive closed expressions for arbitrary 
wavefunction averages in terms of universal coefficients and sums over classical paths,
which contain, besides the supersymmetry results, novel oscillatory contributions. 
Their physical relevance is demonstrated in the context of Coulomb blockade 
physics.
\end{abstract}


%  03.65.Sq  Semiclassical theories and applications.
%  73..20.Fz  Weak or Anderson localization
%  05.45.Mt  Semiclassical chaos (``quantum chaos'').

\maketitle

% \newpage

%\ifpreprintsty\tightenlines \else \begin{multicols}{2} \fi
%\begin{multicols}{2}
%
Since  Chladni's famous experiments on vibrating plates two centuries ago, the morphology of 
eigenfunctions in wave (mechanical) systems has aroused curiosity \cite{stoe}.
In particular, the statistical description of eigenfunctions in quantum (or wave) systems with 
diffusive or chaotic classical (or ray) dynamics has been an intensive research field for 
more than 20 years \cite{his}. Besides the pure academic interest, the study of the 
spatial properties of irregular eigenfunctions provides specific theoretical input 
for experimentally relevant quantities in a wide range of disciplines \cite{stoe}
including optics, acoustics, oceanography \cite{heller}, quantum chaos, and 
mesoscopic condensed matter \cite{bee}.
In the latter, for instance, the energy levels of interacting electrons in quantum dots are 
influenced by wave function correlations entering into interaction matrix elements 
\cite{alhassid,mirlin,denn}.

In the semiclassical limit \cite{gut}, characterized by typical classical actions
 $S_{\rm cl}$ much larger than Planck's constant, 
 i.e.\ $ \hbar_{\rm eff} \equiv \hbar / S_{\rm cl} \ll 1$, 
statistical measures for the spatial structure of eigenfunctions can be deduced from purely
classical quantities, though in a highly non-trivial way. The most spectacular example 
of such a connection is 
embodied in Berry's Random Wave Model (RWM) \cite{berr1}: eigenfunctions of classically chaotic 
 systems possess the same statistical properties as Gaussian random fields with a 
universal spatial two-point correlation function. 

A theory beyond the universal results of the RWM has been rigorously derived  for
disordered systems by an exact map\-ping of the quantum problem onto a supersymmetric 
field theory, the Nonlinear Sigma Model \cite{mirlin}. Semiclassical in spirit, this
Diffusive Sigma Model expresses all results in terms of the classical 
diffusion propagator, and its success has motivated efforts to extend such methods 
to clean chaotic systems. The results of such (still conjectured) 
Ballistic Sigma Models (B$\sigma$M)
\cite{mirlin,ball} are obtained by replacing the diffusion propagator for
the disordered, metallic regime by a suitable ballistic counterpart. Two features of 
the  B$\sigma$M render calculations difficult, already at the level of 
the first non-universal contribution beyond the RWM \cite{mirlinferm}: 
(i) the statistical distribution of eigenfunctions is {\it not Gaussian}, in clear
contrast to the RWM, and
(ii) different eigenfunctions are {\it not independent}.

In this Letter we present an alternative approach to eigenfunction statistics in which 
different eigenfunctions are described as {\it independent Gaussian} fields \cite{note1}.
Employing simple Gaussian integrations, we provide closed expressions for 
general wave function averages for chaotic systems in the semiclassical regime, 
in terms of universal coefficients and sums over classical paths. 
We demonstrate that an arbitrary average contains, besides the universal 
(RWM) result, a system-dependent contribution (e.g.\ from confinement) 
composed of a smooth (diagonal) and an oscillatory part. 
We resolve the apparent contradiction to the supersymmetry techniques 
by showing how the 
results of the B$\sigma$M beyond universality are retained from our result
by neglecting all oscillatory contributions. This finding for chaotic dynamics 
is used to extend our method to the metallic regime of disordered systems, while the physical relevance of the oscillatory terms found is illustrated for
Coulomb Blockade conductance peak statistics.

{\it Defining statistical averages.} Consider a set of $d$ nor\-ma\-lized, real 
solutions $\psi_{n_{\alpha}}$ ($n_{\alpha} \! \in \! \{n_{1},\ldots,n_{d}\}$) of 
the Schr{\"o}dinger equation with non-degenerate, generally non-consecutive eigenvalues 
$E_{n_{1}} < \ldots <E_{n_{d}}$. We study expressions of the general form 
$F(\vec{\psi}_{n_{1}},\ldots,\vec{\psi}_{n_{d}})$
where each entry is a vector
$\vec{\psi}_{n_{\alpha}}=(\psi_{n_{\alpha}}(\vec{r}_{1}^{\alpha}),
\ldots,\psi_{n_{\alpha}}(\vec{r}_{f_{\alpha}}^{\alpha}))$
with $f_{\alpha}$ components depending on different positions 
$\vec{r}_{i}^{\alpha}, i=1,\ldots,f_{\alpha}$ 
(if $\vec{r}_{i}^{\alpha}=\vec{r}^{\beta}_{j}$ for $\alpha \neq \beta$ then $i=j$ by
convention). Upon varying $n_{\alpha}$ inside each window 
$[\bar{n}_{\alpha}-N/2,\bar{n}_{\alpha}+N/2]$,
while keeping all differences $n_{\alpha}-n_{\beta}$ fixed, the function
$F(\vec{\psi}_{n_{1}},\ldots,\vec{\psi}_{n_{d}})$ will exhibit fluctuations. 
The {\it spectral average} (indicated by caligraphic letters) of the function $F$ is 
then naturally defined as
\begin{equation}
\label{eq:prim}
{\cal F}=\frac{1}{N}\sum_{s=-N/2}^{N/2}F(\vec{\psi}_{\bar{n}_{1}+s},
\ldots,\vec{\psi}_{\bar{n}_{d}+s}) \, .
\end{equation}
It is essential to keep in mind that ${\cal F}$ is  fluctuating itself, 
depending on the size and 
location of the energy windows and on the set of positions $\vec{r}_{i}^{\alpha}$.
In the semiclassical limit one has $\bar{n}_{\alpha}\gg N \gg 1$ for all $\alpha$.

Averages  playing  a key role are 
the {\it spatial two-point correlation matrices} ${\bf R}_{\alpha}$ with entries defined as 
\begin{equation}
\label{eq:R}
R_{\alpha}^{i,j}=\frac{1}{N}\sum_{s=-N/2}^{N/2}\psi_{\bar{n}_{\alpha}+s}(\vec{r}_{i}^{\alpha})
\psi_{\bar{n}_{\alpha}+s}(\vec{r}_{j}^{\alpha}) \, .
\end{equation}

{\it Local Gaussian conjecture.} A local Gaussian theory for the spectral averages is based
on two assumptions: (i) $\vec{\psi}_{\bar{n}_{\alpha}-N/2}, \ldots,\vec{\psi}_{\bar{n}_{\alpha}+N/2}$ 
are $N$ realizations of an $f_{\alpha}$-dimensional random vector, denoted by $\vec{v}_{\alpha}$, 
with distribution $P_{\alpha}(\vec{v}_{\alpha})=
(2 \pi)^{-f_{\alpha}/2}\sqrt{{\rm det}{\bf R}_{\alpha}^{-1}}\exp
\left[-\frac{1}{2}\vec{v}_{\alpha}.{\bf R}_{\alpha}^{-1}\vec{v}_{\alpha} \right]$;
(ii) $\vec{v}_{\alpha}$ and $\vec{v}_{\beta}$ are independent random vectors for $\alpha \neq \beta$. 
In the local Gaussian theory all averages can be expressed through
the correlation matrices ${\bf R}_{\alpha}$ by means of
\begin{equation}
\label{eq:F}
{\cal F}^{G}=\int F(\vec{v}_{1},\ldots,\vec{v}_{d})P_{1}(\vec{v}_{1})\ldots
P_{d}(\vec{v}_{d})\, d\vec{v}_{1} \ldots d\vec{v}_{d}  \, .
\end{equation}
The local Gaussian conjecture states that ${\cal F}={\cal F}^{G}$, namely that 
{\it any two eigenfunctions of a classically chaotic system behave like two independent 
Gaussian random fields, each of them uniquely characterized by the exact two-point 
correlation functions}. This is a natural generalization of the ideas presented in \cite{berr1,sied1,yo,kap0} with far reaching consequences when a subsequent  
semiclassical approximation is consistently used.

{\it Semiclassical expansion.} In the semiclassical re\-gime 
each correlation matrix is expanded into a constant part, a leading-order
(in $\hbar_{\rm eff}$) fluctuating part, and higher-order terms \cite{note0}: 
${\bf R}_{\alpha}\!=\!A^{-1}{\bf I}_{\alpha}\!+\! A^{-1}
\tilde{{\bf R}}_{\alpha}\!+\!O(\hbar_{\rm eff}^{3/2})$, 
where ${\bf I_{\alpha}}$ is a $f_{\alpha} 
 \!\times\! f_{\alpha}$ unit matrix and $A$ the system~area for the 2d
case considered. The matrix $A^{-1} {\bf I}_{\alpha}$ 
defines a Gaussian distribution $P^{RMT}(\vec{v}_{\alpha})\!=\!(A/2\pi)^{f_{\alpha}/2}\!\exp 
\left[-(A/2) \vec{v}_{\alpha}\!\cdot\!\vec{v}_{\alpha} \right]$  yielding the 
Random Matrix Theory (RMT) results ${\cal F}^{RMT}$ for the $\alpha$-th state. 
It is straight forward to factorize each probability distribution as
\begin{equation}
\label{eq:dis}
P_{\alpha}(\vec{v}_{\alpha})\!=\!P^{RMT}(\vec{v}_{\alpha})\frac{\exp\!
\left[\frac{A}{2}\vec{v}_{\alpha}\!\cdot\!({\bf I}_{\alpha}\!+\!\tilde{{\bf R}}_{\alpha})^{-1}
\tilde{{\bf R}}_{\alpha}\vec{v}_{\alpha} \right]}{\sqrt{{\rm det}({\bf I}_{\alpha}+
\tilde{{\bf R}}_{\alpha})}}.
\end{equation}
As a key step we note that the semiclassical approach consistently keeps terms up to
second order in the
fluctuating part of the correlation matrices, since the semiclassical approximation neglects 
terms of order $O(\hbar_{\rm eff}^{3/2})$ \cite{gut} 
while $\tilde{{\bf R}}_{\alpha} \sim O(\hbar_{\rm eff}^{1/2})$.
Taylor expansion of the probability distributions in Eq.~(\ref{eq:dis}) to second order 
in ${\tilde {\bf R}}_{\alpha}$ and substitution into Eq.~(\ref{eq:F}) yields
\begin{eqnarray}
\label{eq:finfin}
{\cal F}^{G}&=&{\cal F}^{RMT}+\sum_{\alpha=1}^{d}
    \sum_{i,j=1}^{f_\alpha} \tilde{R}_{\alpha}^{i,j} \times  \\
& & \times \left[
    {\cal F}_{i,j}^{\alpha} +
\sum_{k,l=1}^{f_{\alpha}}{\cal F}_{i,j,k,l}^{\alpha}
\tilde{R}_{\alpha}^{k,l}  
+ 
\sum_{\beta < \alpha}^{d}\sum_{k,l=1}^{f_{\beta}}
{\cal F}_{i,j,k,l}^{\alpha,\beta}\tilde{R}_{\beta}^{k,l}\right] 
\nonumber 
\end{eqnarray}
with universal (system-independent) coefficients 
\begin{eqnarray}
\label{eq:fs}
{\cal F}^{RMT}&=&\langle F(\vec{v}_{1},\ldots,\vec{v}_{d})\rangle^{RMT}, \nonumber\\ 
{\cal F}_{i,j}^{\alpha}&=&\langle F(\vec{v}_{1},\ldots,\vec{v}_{d})q_{i,j}
(\vec{v}_{\alpha})\rangle^{RMT}, \nonumber\\
{\cal F}_{i,j,k,l}^{\alpha}&=&\langle F(\vec{v}_{1},\ldots,\vec{v}_{d})
q_{i,j,k,l}(\vec{v}_{\alpha}) \rangle^{RMT}\, , \\
{\cal F}_{i,j,k,l}^{\alpha,\beta}&=&\langle F(\vec{v}_{1},\ldots,\vec{v}_{d})
q_{i,j}(\vec{v}_{\alpha})q_{k,l}(\vec{v}_{\beta})\rangle^{RMT} \, .\nonumber
\end{eqnarray}
Here $\langle\ldots\rangle^{RMT}$ denotes an average with respect to the distribution 
$\prod_{\eta=1}^{d}P^{RMT}(\vec{v}_{\eta})$, and $q(\vec{v})$ are simple polynomials 
of the components of its argument $\vec{v}$:
\begin{eqnarray}
\label{eq:qs}
q_{i,j}(\vec{v})&=&\frac{1}{2}(Av_{i}v_{j}-\delta_{ij}) \; , \nonumber \\
q_{i,j,k,l}(\vec{v})&=&\frac{1}{2}[q_{i,j}(\vec{v})q_{k,l}(\vec{v})+
        2q_{i,k}(\vec{v})q_{j,l}(\vec{v})] \\ 
        & &-\frac{1}{4}A^2v_iv_jv_kv_l \, . \nonumber
\end{eqnarray}

{\it Semiclassical correlation function}. The fluctuating part of the correlation matrix 
is obtained from the semiclassical Green function and is given by \cite{sied1,yo}
\begin{eqnarray}
\label{eq:sc}
\tilde{R}_{\alpha}^{i,j}&=&-\delta_{i,j}+J_{0}\left(k_{n_{\alpha}}|\vec{r}^{\alpha}_{i}-
  \vec{r}^{\alpha}_{j}| \right) + \\ 
  &+&\left(\frac{2 \hbar}{m^{2} \pi}\right)^{1/2} \sum_{\gamma_{i,j}}\Gamma\left(
  \frac{T_{\gamma_{i,j}}}{\tau_{N}}\right)
\left|D_{\gamma_{i,j}}\right| ^{1/2}\cos\left(\frac{S_{\gamma_{i,j}}}{\hbar}\right) \nonumber 
\end{eqnarray}
as a sum over classical trajectories  $\gamma_{i,j}$ joining $\vec{r}^{\alpha}_{i}$ with 
$\vec{r}^{\alpha}_{j}$ at fixed energy $E_{n_\alpha} \simeq \bar{n}_{\alpha}\Delta$, 
where $\Delta$ denotes the (constant) mean level spacing.
The Bessel function $J_{0}(x)$ with $k_{n_{\alpha}}\!=\!\sqrt{2mE_{n_{\alpha}}}/ \hbar$ 
is the contribution from the unique direct trajectory between $\vec{r}^{\alpha}_{i}$ and 
$\vec{r}^{\alpha}_{j}$. The sum in Eq.~(\ref{eq:sc}) is taken over all non-direct orbits $\gamma_{i,j}$
with actions $S_{\gamma_{i,j}}(\vec{r}_{i}^{\alpha},\vec{r}_{j}^{\alpha},
E_{n_{\alpha}})=\int_{\gamma_{i,j}} \vec{p}\cdot d\vec{r}$ and semiclassical prefactors
$D_{\gamma_{i,j}}$ where the stability and topology of each path enters \cite{gut}. 
The window function $\Gamma(x)=\sin(x)/x$ suppresses contributions from trajectories
with traversal time $T_{\gamma_{i,j}}=\partial S_{\gamma_{i,j}}/\partial E_{n_{\alpha}}$ 
larger than the characteristic time $\tau_{N}=2 \hbar / N \Delta$ related to the energy
average (\ref{eq:prim}). 

Equation (\ref{eq:finfin}), supplemented by the definitions (\ref{eq:fs},\ref{eq:qs}) and 
Eq.~(\ref{eq:sc}) for $\tilde{R}_{\alpha}^{i,j}$, is our main result. We illustrate its
power by computing the intensity distribution
$I(\psi(\vec{r}_{1}))=\delta(t-A\psi(\vec{r}_{1})^{2})$, a prominent and frequently studied 
\cite{his,mirlin} measure for wave function statistics.
While the evaluation based on supersymmetry methods is quite involved {\cite{mirlinferm}, 
using our Gaussian approach (with $d=1, f_{1}=1$)
we trivially find for the universal polynomials (\ref{eq:qs}) 
$q_{1,1}(v)=\frac{1}{2}(Av^{2}-1)$ and 
$q_{1,1,1,1}(v)=\frac{3}{8}(Av^{2}-1)^{2}-\frac{1}{4}A^{2}v^{4}$. 
The  coefficients (\ref{eq:fs}) are then obtained without any further integration,
and Eq.~(\ref{eq:finfin}) yields the intensity distribution
\begin{equation}
\label{eq:pt} 
{\cal I}^{G}(t)=\frac{e^{-\frac{t}{2}}}{\sqrt{2 \pi t}}
\left[1\!+\!\frac{\tilde{R}^{1,1}}{2}(t\!-\!1)\!+\!\frac{(\tilde{R}^{1,1})^{2}}{4}
(3\!-\!6t\!+\!t^{2})\right] 
\end{equation}
in terms of closed orbits (through $\tilde{R}^{1,1}$) starting and ending at $\vec{r}_{1}$. 
Equation~(\ref{eq:pt}) includes both the universal limit $e^{-\frac{t}{2}}/ \sqrt{2 \pi t}$
(the Porter-Thomas distribution \cite{bee}) and the sigma 
model result of \cite{mirlinferm} as we will show below.

{\it Diagonal approximation and sigma model}. For energy windows of size 
$N_{{\rm Th}} > \sqrt{A}k_{n_{\alpha}}$, corresponding to $\tau_N$ in Eq..~(\ref{eq:sc})
smaller than the time of flight through the system, i.e.\ the ballistic Thouless time 
$\tau_{\rm Th}$,
all contributions beyond the direct path are damped out. 
In this universal regime the RWM predictions (consistent with RMT) are given by substitution of the direct path contribution to $\tilde{R}_{i,j}$ (first line in Eq.~(\ref{eq:sc})) into Eqs.~(\ref{eq:dis}) or (\ref{eq:finfin}). It has been shown that the Gaussian and sigma-model results coincide at this universal level \cite{mirlinferm,sied2}. For smaller energy 
windows, however, system-dependent deviations appear, i.e., 
${\cal F}={\cal F}^{RWM}+{\cal F}^{SYS}$. In the present Gaussian
approach deviations ${\cal F}^{G,SYS}$ from universality are obtained by substitution of 
the non-direct contribution to $\tilde{R}_{i,j}$ (second line of Eq.~(\ref{eq:sc})) 
into Eq.~(\ref{eq:finfin}). As we see, ${\cal F}^{G,SYS}$ consists of coherent single and 
double sums over non-direct classical paths 
of increasing length, while deviations ${\cal F}^{\sigma,SYS}$ from universality 
in the B$\sigma$M are expressed through a purely classical object, the ballistic propagator \cite{mirlinferm}. 
It is by no means clear whether the two approaches for ${\cal F}^{SYS}$ are consistent, 
and we address this fundamental issue now. 

The frequency-dependent ballistic propagator $\Pi^{i,j}(w)$ is constructed by projecting 
the resolvent of the classical Liouville equation, $\{H(\vec{r},\vec{p}),\rho\}=iw\rho$ 
(where $\{\ldots \}$ is the Poisson bracket), onto configuration space at energy 
$E_{n_{\alpha}}=p_{n_{\alpha}}^{2}/2m$. The first step in unifying the local Gaussian 
and sigma-model approaches consists in expressing $\Pi^{i,j}(w)$ through a sum over 
non-zero paths \cite{mar}, $\Pi^{i,j}(w)=\Pi^{i,j}_{0}(w)+\sum_{\gamma_{i,j}}D_{\gamma_{i,j}}e^{iwT_{\gamma_{i,j}}}$, where the contribution $\Pi_{0}$ from direct paths (set to zero when $\vec{r}_{i}=\vec{r}_{j}$) contains $D^{0}_{i,j}=m^{2}/p_{n_{\alpha}}|
\vec{r}_{i}-\vec{r}_{j}|$ and $T^{0}_{i,j}=m|\vec{r}_{i}-\vec{r}_{j}|/p_{n_{\alpha}}$. 
The related energy-averaged, smoothed version reads
\begin{equation}
\label{eq:pi}
\tilde{\Pi}^{i,j}(w)=\Pi_{0}^{i,j}(w)+
\sum_{\gamma_{i,j}}D_{\gamma_{i,j}}\Gamma^{2}\left(\frac{T_{\gamma_{i,j}}}{\tau_{N}}\right)
e^{iwT_{\gamma_{i,j}}} 
\end{equation}
with $\tilde{\Pi}^{i,j}(w)\simeq \Pi^{i,j}(w)$ for $\tau_N \gg \tau_{\rm Th}$. 
Consider now terms in the double sums in Eq.\ (\ref{eq:finfin}) where the classical 
paths involved join the same points, namely $\vec{r}_{i}^{\alpha}\! =\!
\vec{r}_{i}^{\beta}\!=\!\vec{r}_{i}$ and $\vec{r}_{j}^{\alpha}\!=\!\vec{r}_{j}^{\beta}
\!=\!\vec{r}_{j}$. The expression $\tilde{R}^{i,j}_{\alpha}\tilde{R}^{i,j}_{\beta}$ is a 
double sum over terms oscillatory in $[S_{\gamma_{i,j}}(E_{\bar{n}_{\alpha}}) 
\pm S_{\gamma_{i,j}}(E_{\bar{n}_{\beta}})]/\hbar$. When the energy difference 
$\hbar w_{\alpha,\beta}=\Delta (n_{\beta}-n_{\alpha})$ is classically small, 
each trajectory in $\tilde{R}^{i,j}_{\beta}$ is a smooth deformation of a corresponding 
trajectory in $\tilde{R}^{i,j}_{\alpha}$, and we can expand 
$S_{\gamma_{i,j}}(E_{\bar{n}_{\beta}})\simeq S_{\gamma_{i,j}}(E_{\bar{n}_{\alpha}})+\hbar  
w_{\alpha,\beta} T_{\gamma_{i,j}}$. 
Collecting the non-oscillatory terms in the action 
difference of non-direct paths we obtain the so-called diagonal part
\begin{equation}
\label{eq:diag}
\left[\tilde{R}^{i,j}_{\alpha}\tilde{R}^{i,j}_{\beta} \right]_{\rm diag}^{SYS}=\frac{2 \hbar}{m^{2} \pi} {\rm Re}\sum_{\gamma_{i,j}}D_{\gamma_{i,j}}\Gamma^{2}\left(\frac{T_{\gamma_{i,j}}}{\tau_{N}}\right)
e^{iw_{\alpha,\beta}T_{\gamma_{i,j}}}  \, .
\end{equation}
Substitution of Eqs.\ (\ref{eq:pi},\ref{eq:diag}) 
in Eq.~(\ref{eq:finfin}) shows that the diagonal part of 
${\cal F}^{G, SYS}$ is of order $\hbar_{\rm eff}$ and given by a linear combination 
of smoothed ballistic propagators (with universal coefficients 
${\cal F}_{i,j,k,l}^{\alpha},{\cal F}_{i,j,k,l}^{\alpha,\beta}$) 
with the direct path contribution excluded. 
Using the general formulas (\ref{eq:fs},\ref{eq:qs}) the diagonal contribution 
to ${\cal F}^{G,SYS}$ is readily calculated, giving in the limit $\tau_{N} \gg \tau_{\rm Th}$ 
exactly the various specific B$\sigma$M results\cite{note2} available in the literature
(moments of the wavefunction, distribution of intensities, 
two-energies four-point correlations \cite{mirlin}, and two-point 
intensity distributions \cite{prig}). Hence we
conclude that {\em to lea\-ding order in the deviation from universality, 
the B$\sigma$M corresponds to a Gaussian theory in diagonal approximation}. 
For instance, Eq.\ (\ref{eq:pt}) yields 
$\hbar/(2 \pi m^2) [\exp(-t/2)/ \sqrt{2 \pi t}](3-6t+t^2) $ 
for the prefactor of the ballistic propagator in the intensity distribution 
${\cal I}^{G}(t)$,
in perfect agreement with supersymmetry \cite{mirlinferm}.

Here several remarks are due: 
(i) Possibly most impor\-tant\-ly, our method provides also the general 
leading-order deviation from universality for the body of averages in disordered 
systems in the metallic regime by simply replacing $\tilde{\Pi}^{i,j}(w)$ by 
the diffusive propagator;
(ii) there is not a single chaotic system where the exact ballistic propagator
$\Pi^{i,j}(w)$ is known; hence the use of few classical paths to 
construct the smoothed version (\ref{eq:pi}) makes the Gaussian theory more 
accessible for practical calculations;
(iii) the direct path contribution, known to be counted twice as an artifact
in the sigma model calculations \cite{mirlinferm}, is correctly incorporated in 
the present approach;
(iv) our results can be easily generalized to the case of broken time-reversal symmetry 
(by taking each $\vec{v}_{\alpha}$ in Eq.~(\ref{eq:F}) as a complex vector with independent 
real and imaginary parts) and to the case of smooth potentials by using a sum
over classical paths instead of Eq.~(\ref{eq:sc}).

{\it Statistics and oscillations of Coulomb Blockade transmission peaks.}
As  a further application of our approach we consider transport through a quantum dot 
weakly coupled to two leads. In this Coulomb Blockade regime, characterized by 
the mean resonance width $\bar{\Gamma} \ll k_{\rm B}T \ll \Delta$,
transport is mediated by resonant tunneling with corresponding distinct conductance 
peaks. These Coulomb Blockade peaks and the fluctuations of their heights have been
prominent objects of experimental studies \cite{CB}. 
The universal contribution to the conductance distribution, derived in 
RMT \cite{CB1} for the case of one-channel leads, was extended to the multichannel 
case for ballistic quantum dots using the RWM \cite{CB2} 
and for disordered systems using the sigma model \cite{CB3}, while effects due to 
periodic orbits where studied in \cite{kap}. 
However, for disordered dots in the Coulomb Blockade regime we are not aware of 
any prediction beyond universality. Moreover, non-universal oscillatory effects 
in ballistic dots were presented in \cite{CB4}, 
but the first-order theory used there 
fails to reproduce numerical results for the case of asymmetric leads \cite{nar1}. 

Here we apply our approach to these two problems. To this end
we consider leads supporting one channel each, connected to the dot at 
positions $\vec{r}_{i}$ ($i=1,2$) with equal coupling strength $\alpha$, 
generalizations to more channels are straight forward \cite{RDU-KR}.
Following \cite{CB1,CB2}, the height of the $n_{\alpha}$-th conductance peak is 
given by $G_{n_{\alpha}}=(e^{2}/h)(\alpha \Delta)/(2\pi k_{\rm B}T)g_{n_\alpha}$ 
where 
\begin{equation}
g_{n_{\alpha}}=\pi A\frac{\psi^{2}_{n_{\alpha}}(\vec{r}_{1})\psi^{2}_{n_{\alpha}}
(\vec{r}_{2})}{\psi^{2}_{n_{\alpha}}(\vec{r}_{1})+\psi^{2}_{n_{\alpha}}(\vec{r}_{2})},
\end{equation} 
which fluctuates with $n_{\alpha}$. Applying our general Eq.\
(\ref{eq:finfin}) we obtain the distribution of conductances as 
$P(g)=(2 \pi g)^{-1/2}{\rm exp}
\left[-g/2\right]\left[1+\delta P^{(1)}(g)+\delta P^{(2)}(g) \right]$ with first 
and second order deviations from universality given by 
\begin{eqnarray}
\delta P^{(1)}(g)&= &q^{(1)}(g)\left[\tilde{R}^{1,1}+\tilde{R}^{2,2}\right] 
\, , \nonumber \\
\delta P^{(2)}(g)&=
&q_{1}^{2}(g)\left[(\tilde{R}^{1,1})^{2}+(\tilde{R}^{2,2})^{2}
                     \right] + \\
&+& q_{2}^{(2)}(g)\tilde{R}^{1,1}\tilde{R}^{2,2}+q_{3}^{(2)}(g)(\tilde{R}^{1,2})^{2}\, .
                                \nonumber
\end{eqnarray}
Here the polynomial corrections to the RMT result,
$q^{(1)}(x)(1/4)(x\!-\!1),q_{1}^{(2)}(x)=(1/32)(x^{2}\!-\!9x\!+\!6),q_{2}^{(2)}(x)=
(1/16)(x^{2}\!-\!3x),$ and $q_{3}^{(2)}=(1/8)(x^{2}\!-\!7x\!+\!4)$,
are modulated by the single and double sums over classical paths.

Using $P(g)$, we now calculate the mean of the conductance peak heights,
$\langle g \rangle =1+\delta g^{(1)}+\delta g^{(2)}$. 
First we discuss the leading term 
$\delta g^{(1)}=(1/2)(\tilde{R}^{1,1}+\tilde{R}^{2,2})\sim 
O(\hbar_{{\rm eff}}^{1/2})$. In a billiard system with energy window
$\delta e \simeq \hbar/ \tau_{\rm Th}$ we have $\tilde{R}^{i,i} \sim \cos (kL_{i})$ where $L_{i}$ is the length of the 
shortest classical trajectory starting and ending at $\vec{r}_{i}$. 
The corresponding modulations of the conductance   
with frequencies $L_{1},L_{2}$ were already reported in \cite{nar1,CB4}.

Our approach now enables us to go beyond the leading semiclassical order and 
to provide the explicit expression 
$\delta g^{(2)}=-(3/16)(\tilde{R}^{1,1}-\tilde{R}^{2,2})^{2}-(1/4)(\tilde{R}^{1,2})^{2}\sim O(\hbar_{{\rm eff}})$. As we see, $\delta g^{(2)}$ 
will be a combination of terms $\cos[k(L_{i}\pm L_{j})]$. We conclude that the second-order 
calculation is essential to understand the modulations with frequency 
$L_{1}\pm L_{2}$ numerically observed for asymmetric leads  
in Ref.\ \cite{nar1}. To our knowledge neither the first- nor the 
second-order oscillatory effect, present in Coulomb blockade physics, is accessible using 
the B$\sigma$M.

For diffusive dots in the metallic regime $\delta g^{(1)}$ is exponentially supressed and we just replace in the diagonal contribution to 
$\delta g^{(2)}$ the smoothed ballistic propagator  $\tilde{\Pi}^{i,j}(w)$
by its diffusive counterpart $\Pi_{\rm dis}^{i,j}(w)$ to obtain
\begin{equation}
\delta g^{{\rm dis}}=-\frac{3\hbar}{8\pi m^{2}}[\Pi_{\rm dis}^{1,1}(0)+\Pi_{\rm dis}^{2,2}(0)
+4\Pi_{\rm dis}^{1,2}(0)] \, .
\end{equation}
Since $\Pi_{\rm dis}^{i,j}(0) >0$, we predict that the leading non-universal correction 
to the mean conductance peak height in weakly disordered quantum dots (with single-channel 
point contacts) is {\em always negative} and of order $\hbar_{\rm eff}$.

To summarize, we have used a local Gaussian theory for eigenfunction statistics 
to derive both smooth and oscillatory effects beyond (and including) the universal 
Random Wave Model. Smooth contributions are shown to give existing results of the 
Sigma Model, illuminating the connection between the two methods beyond 
universality. In view of that, we use our approach to present new 
results for the conductance of diffusive quantum dots in the Coulomb Blockade 
regime. For the ballistic case, new oscillatory effects neglected both by the 
Sigma Model and previous semiclassical approaches are shown to describe 
previously unexplained numerical results.   

{\it Acknowledgements.} 
We gratefully acknowledge conversations with C.\ Lewenkopf and A.\ Mirlin. 
This work was supported by the {\em Deutsche Forschungsgemeinschaft} (research school GRK 638). 
JDU acknowledges additional support from the {\em Minerva Center for Nonlinear Physics of Complex Systems} at the Weizmann Institute of Science
(through GIF grant 808/2003) where this work was completed.
KR thanks MPI-PKS Dresden for the kind hospitality during the final stage of the work.

%%%%%%%%%%%%%%%%%%%%%%%%%%%%%%%%%%%%%%%%%%%%%%%%%%%%%%%%%%%%%%%%%%%%%%%%%%%%%
%
%  Bibliography
%


\begin{thebibliography}{10}
\bibitem{note1}
Correlations owing to a proper orthonormalization can be incorporated into
our approach \cite{RDU-KR}.
\bibitem{note2}
No general expression as Eq.~(\ref{eq:finfin}) has been derived using the sigma model.
\end{thebibliography}
\end{document}